\documentclass[12pt]{article}
\usepackage{graphicx}
\begin{document}

\begin{center} 
CYCLE-TO-CYCLE FLUCTUATIONS OF BURNED \\ FUEL MASS IN SPARK IGNITION 
COMBUSTION 
ENGINES
\vspace{1cm}

{\bf Miros\l{}aw  Wendeker}$^1$, {\bf Grzegorz 
Litak}$^2$,
{\bf Marcin Krupa}$^1$
\vspace{0.5cm}

$^1$Department of Internal Combustion Engines, \\
Technical University of Lublin \\ Nabystrzycka 36,
PL-20-618 Lublin, Poland \\
$^2$Department of Applied Mechanics, \\
Technical University of Lublin \\ Nabystrzycka 36,
PL-20-618 Lublin, Poland
\end{center}
\vspace{1cm}

\abstract{We examine a simple, fuel-air, model of 
combustion
in spark ignition (si) engine with indirect injection.
In our two fluid model, variations of fuel mass burned in cycle 
sequences appear
due to stochastic 
fluctuations of a fuel feed amount. We have shown 
that a small amplitude of these fluctuations affects  
considerably the stability of a 
combustion process strongly depending on the quality of air-fulel mixture. 
The largest influence was found in the limit of a lean combustion.   
The possible effect of nonlinearities in the 
combustion process were also 
discussed.
}
\vspace{0.5cm}



\section{Introduction}

\noindent Cyclic combustion variability,  found in 19th century 
by Clerk (1886) in
all spark
ignition (si) engines, has attracted great interest of researchers during 
last 
years (Heywood 1988, Hu 1996, Daw {\em et al.} 1996, 1998, 2000, Wendeker
{\em et al.} 2003, 2004). 
Its elimination 
 would give 10\%
increase
in the power output  of the engine 
(Heywood 1988).  
The key source of their existence may be associated with either 
stochastic disturbances 
(Roberts {\em et al.} 1997,
Wendeker {\em et al.} 1999) 
or 
nonlinear dynamics   
(Daw {\em et al} 1996, 1998)
 of 
the combustion process. 
Daw {\it et al.} 
(1996, 1998)
  and more recently
Wendeker {\it et al.} 
(2003, 2004)
have done  the nonlinear analysis of the experimental 
data of such a process.
Changing an advance spark angle  they observed  the
considerable increase of the noise level
(Wendeker {\it et al.} 2003) claiming that it is due to 
chaotic
dynamics of the process.
On the other hand the main sources of cyclic variability were  
classified by Heywood 
(1988) as
the aerodynamics in the cylinder during
combustion, the amount of fuel, air and recycled exhaust
gases supplied to the cylinder and a
mixture composition near the spark plug.
In this paper we will model the variation of fuel ignition amount as the 
most 
common source of instability in indirect injection.

The present paper is organized as follows. After the introduction in the 
present 
section we define the model by a set of difference equations in the next 
section (Sec. 2). 
 This model, in deterministic and stochastic forms, will be applied in 
Sec. 
3, where we analyze the oscillations of burned mass. Finally we 
derive conclusions and remarks in Sec. 4.  

\section{Two fluid model of fuel-air mixture combustion}

\noindent Starting from fuel-air mixture
we define the time evolution of the corresponding amounts.
Namely, we will follow
the time histories of the masses of fuel $m_f$, and air $m_a$.

\begin{table}
\caption{\label{tableone} Constants and variables of the model.
}
\vspace{1cm}
\hspace{2cm}
\begin{tabular}{c|c}
 \hline
stoichiometric coefficient  & $s=14.63$
 \\
exhaust ratio    & $\alpha=0.92$  \\
air mass in a cylinder  & $m_a$    \\
fuel mass in a cylinder  & $m_f$  \\
fresh air amount & $\delta m_a$   \\
fresh fuel amount & $\delta m_f$  \\
air/fuel ratio & $r=m_a/m_f$  \\
burned fuel mass & $\Delta m_f$  \\
combusted air mass & $\Delta m_a$  \\
air/fuel equivalence ratio & $\lambda$  \\ 
random number generator & $ N(0,1,i)$  \\
mean value \\ of fresh fuel amount & $\delta m_{fo}$  \\
standard deviation \\of fresh fuel amount & $\sigma_{mf}$  \\
standard deviation \\ of  the equivalence ratio & $\sigma_{\lambda}$  \\
\hline
\end{tabular}
\vspace{1cm}
\end{table}

\begin{figure}
\begin{center}
\includegraphics[scale=0.4,angle=-90]{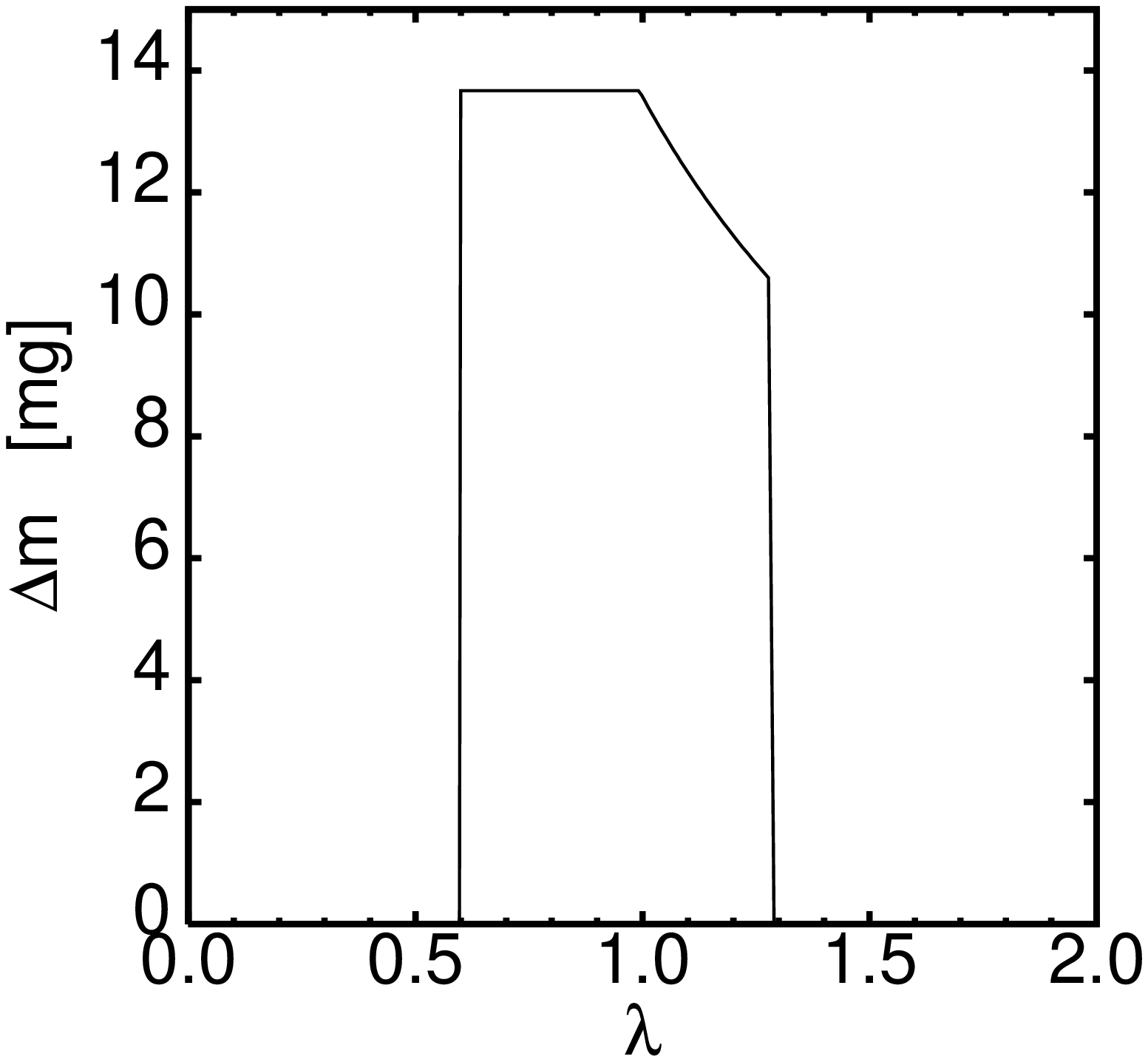}
\caption{The combustion curve $\Delta m (\lambda)$ for the constant  fresh air feed
$\delta 
m_a=200$ mg.}
\end{center}
\end{figure}

Firstly, we assume the initial value of  $m_a(i)$, $m_f(i)$ 
automatically their ratio
$r(i)$:
\begin{equation}
r(i) =\frac{m_a(i)}{m_f(i)}
\end{equation}
for  $i=0$.

Secondly, depending on parameter $r$ with reference to a stoichiometric 
constant $s$ we have two possible cases:
 fuel and air deficit, respectively.
For a deterministic model, the first case lead to  
\begin{equation}
r(i) > 1/s
\end{equation}
we calculate next masses using following difference equations:
\begin{eqnarray}
m_f (i+1) &=& (1-\alpha) \left(m_f(i) -\frac{1}{s}
m_a (i) \right) + \delta m_f
\nonumber \\
m_a (i+1) &=&  \delta m_f,
\end{eqnarray}
where $\alpha$ is the exhaust ratio of the engine,
 $\delta m_f$ and  $\delta m_a$ denotes fresh fuel and air amounts added 
in each
combustion cycle $i$. 
In the opposite (to Eq. 2.2)  case
\begin{equation}
r(i) < 1/s
\end{equation}
we use the different formula
\begin{eqnarray}
m_f (i+1) &=&  \delta m_f
\nonumber \\
m_a (i+1) &=& (1-\alpha) \left(m_a(i)- s m_f(i) \right) + \delta m_a
\end{eqnarray}

Note that variables $m_a$ and $m_f$ are the minimal set of our interest.
From the above equations  one can easily calculate other interesting 
quantities as the combusted masses of 
fuel $\Delta m_f(i)$ and air $\Delta m_a(i)$ and air-fuel equivalence 
ratio before each combustion event $i$:
\begin{equation}
\lambda \approx s\frac{m_a(i)+(1-\alpha)\Delta 
m_a(i-1)}{m_f(i)+(1-\alpha)\Delta 
m_f(i-1)}  
\end{equation}
Basing on experimental results we use the additional necessary condition 
(Kowalewicz 1984)
of
combustion process
\begin{equation}
0.6 < \lambda < 1.3 . 
\end{equation}
For better clarity our notations of system parameters: constants and 
variables are 
summarized in Tab. 1.

Basing on the relations (Eqs. 2.1-2.7) we plotted the combustion 
curve for the assumed constant fresh air feed $\delta m_a=200$ mg.

\begin{figure}
 \begin{center}
\includegraphics[scale=0.3,angle=-90]{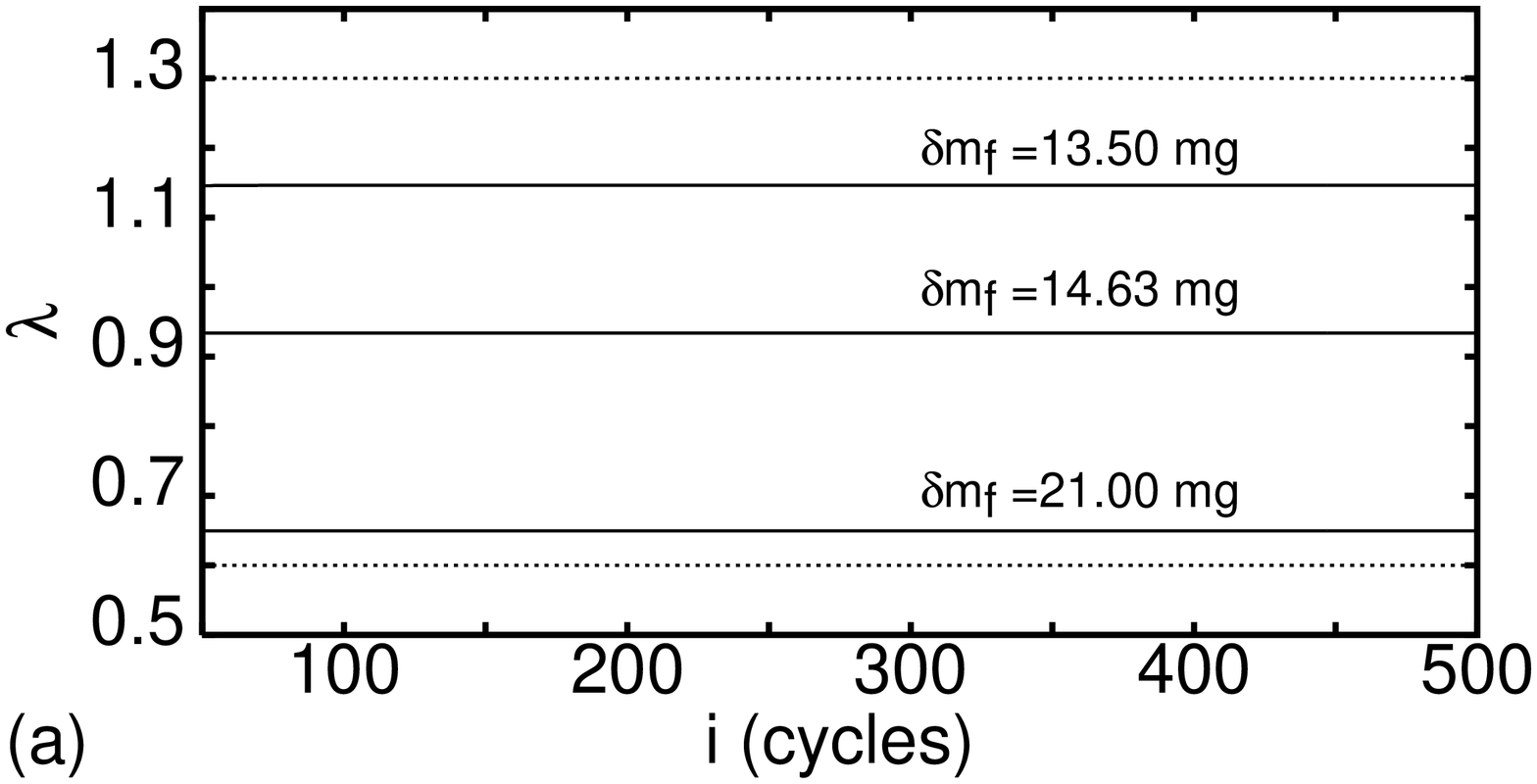}
\includegraphics[scale=0.3,angle=-90]{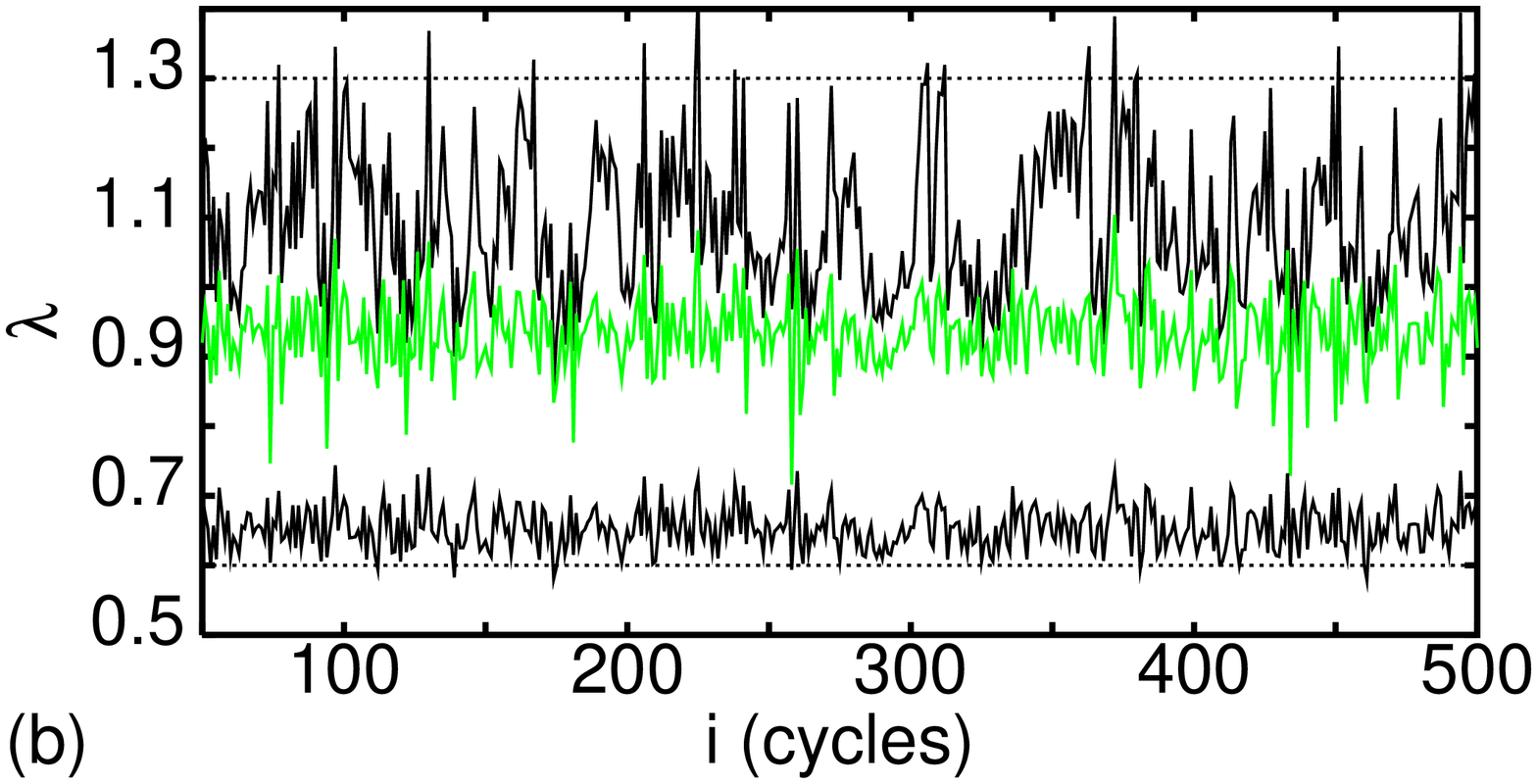}
\caption{The dependence of $\lambda$ on sequential cycles $i$ for deterministic (a) and 
stochastic (b) processes. The fresh 
air amount was assumed to be $\delta m_a=200$ mg while the fresh
fuel amount varies: $\delta m_f=13.50$ mg , 14.63 mg, 21.00 mg starting from the top
curve,
respectively.}
 \end{center}
\end{figure}

Finally, in the case of stochastic injection,  instead of 
constant $\delta m_f(i)$ (Eqs. 2.3 and 2.5) (for each cycle $i$) we 
introduce
its mean value  $\delta m_{fo}={\rm const.}$, while $\delta m_f$ in the 
following way:
\begin{equation}
\delta m_f(i)=\delta m_{fo}+ \sigma_{mf}N(0,1,i),
\end{equation}
where $N(0,1,i)$ represents random number generator giving a sequence $i$ 
of numbers with  a 
unit-standard deviation of normal 
(Gaussian) 
distribution and the nodal mean. The scaling factor $\sigma_{mf}$ 
corresponds to the 
mean standard deviation of the fuel injection amount.
The cyclic variation of $\delta m_f(i)$ can be associated with such 
phenomena as  fuel vaporization and  fuel-injector variations.

\section{Oscillations of burned fuel mass}

\noindent Here we describe the results of simulations. Using Eqs. 2.1-2.8
we have performed recursive calculations for deterministic and 
stochastic conditions and obtained time histories of 
various system parameters:
$m_f$, $m_a$, $\Delta m_f$, $\Delta m_a$ and $\lambda$.
The results for $\lambda$ are shown in Fig. 2. The upper panel (Fig. 2a), 
corresponding to deterministic combustion for three different values of  
fuel 
injection parameter $\delta m_{f}$, 
shows $\lambda$ as straight lines versus cycle $i$, while the lower 
(Fig. 2b) one
reflects the variations of $\lambda$ in stochastic conditions. 
The order of curves appearing in the Fig. 2b  is the same as in Fig. 2a
stating from the smallest value of considered fuel injection amounts from 
the top.  
In stochastic simulations we used input of random fuel injection with 
standard deviation 
equal to 10\% of its mean value
$\sigma_{mf}= 0.1~ \delta m_{fo}$. 
The obtained results clearly indicate that the  fluctuations of $\lambda$
are growing with larger $\lambda$. This can be also found by analytical 
evaluation of Eq. 2.6. It is not difficult to check that
\begin{equation}
\sigma_{\lambda} \sim \lambda^2 \sigma_{mf}.
\end{equation}

\begin{figure}
 \begin{center}
\includegraphics[scale=0.3,angle=-90]{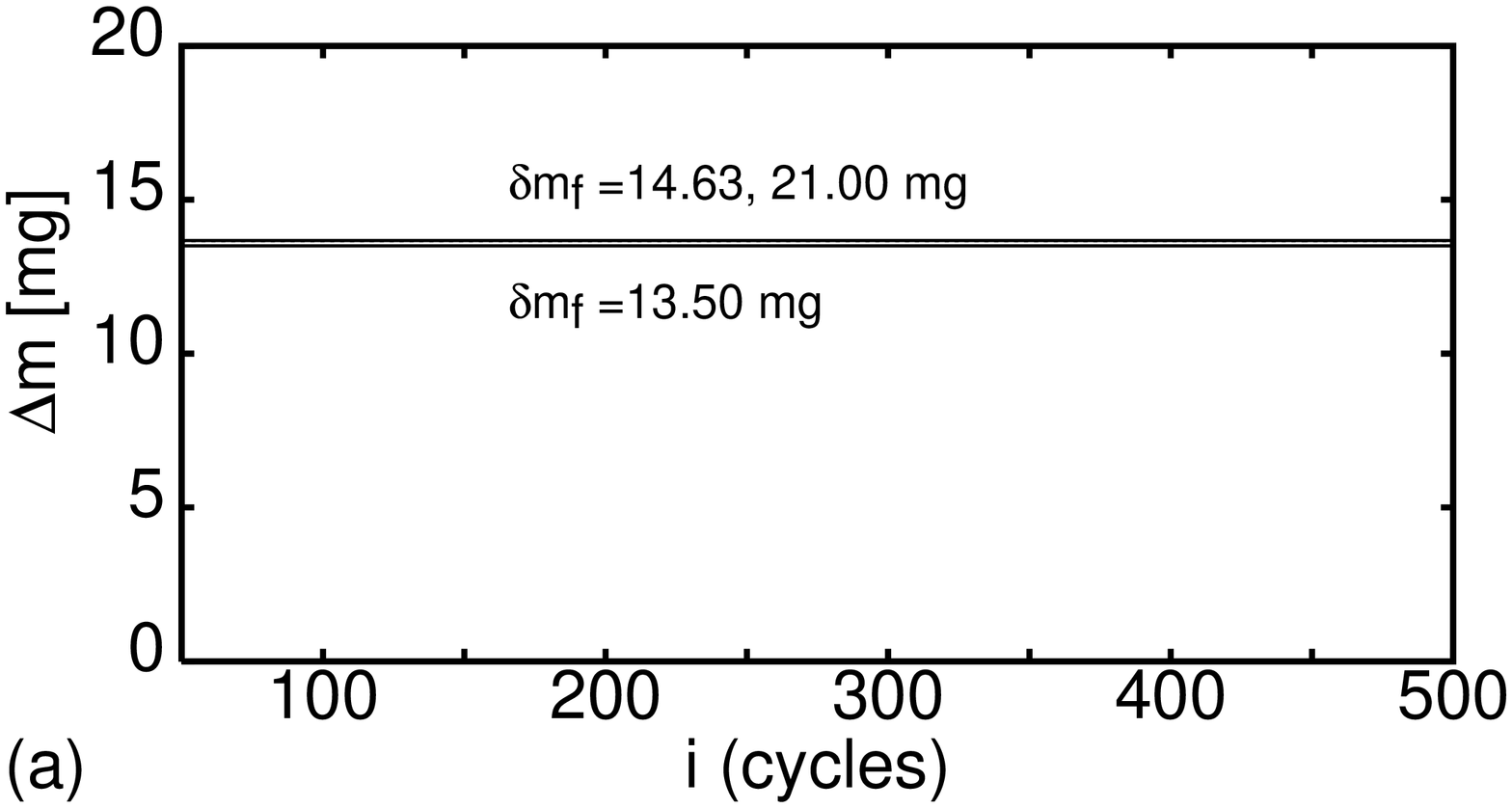}
\includegraphics[scale=0.3,angle=-90]{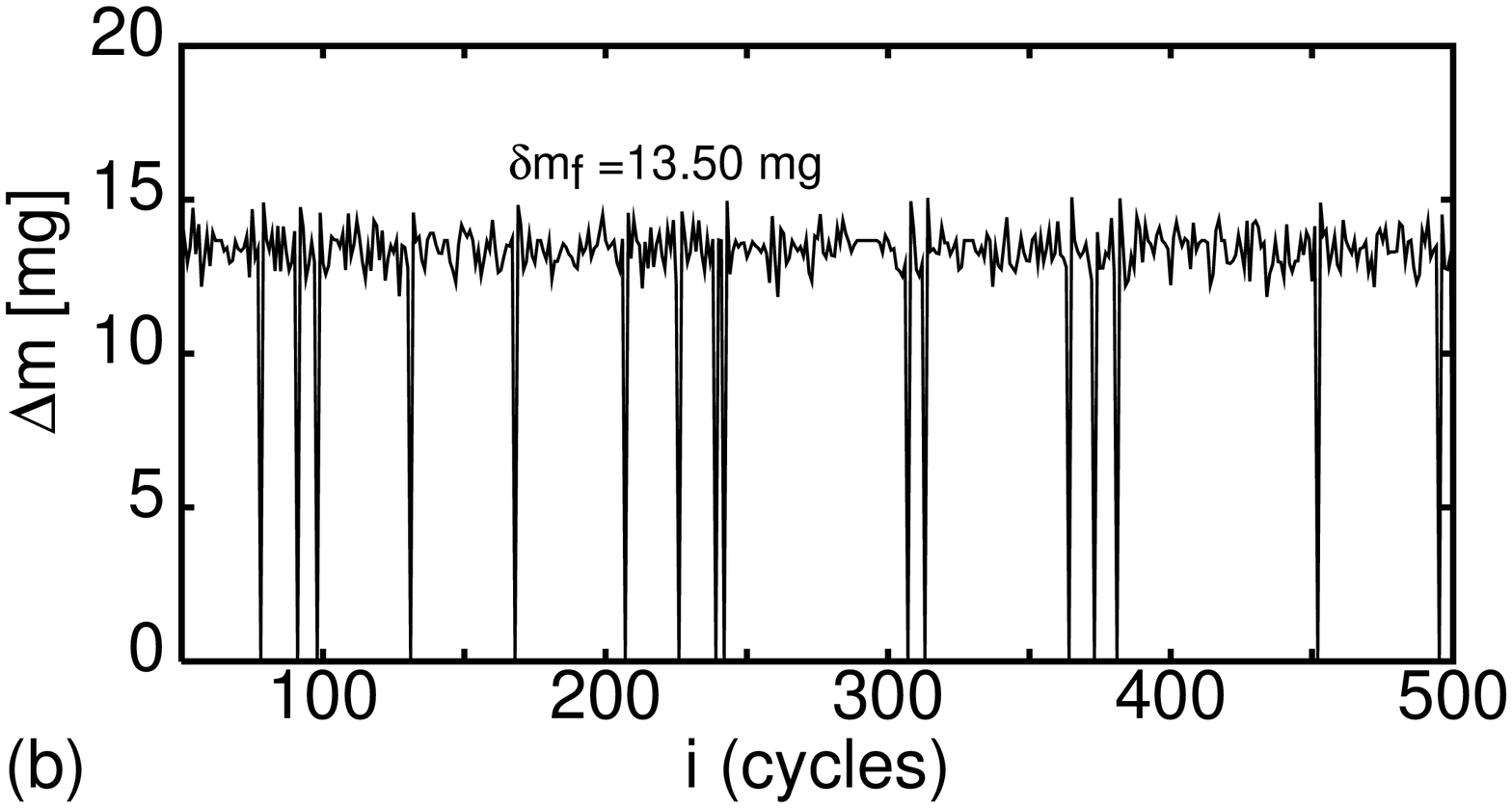}
\includegraphics[scale=0.3,angle=-90]{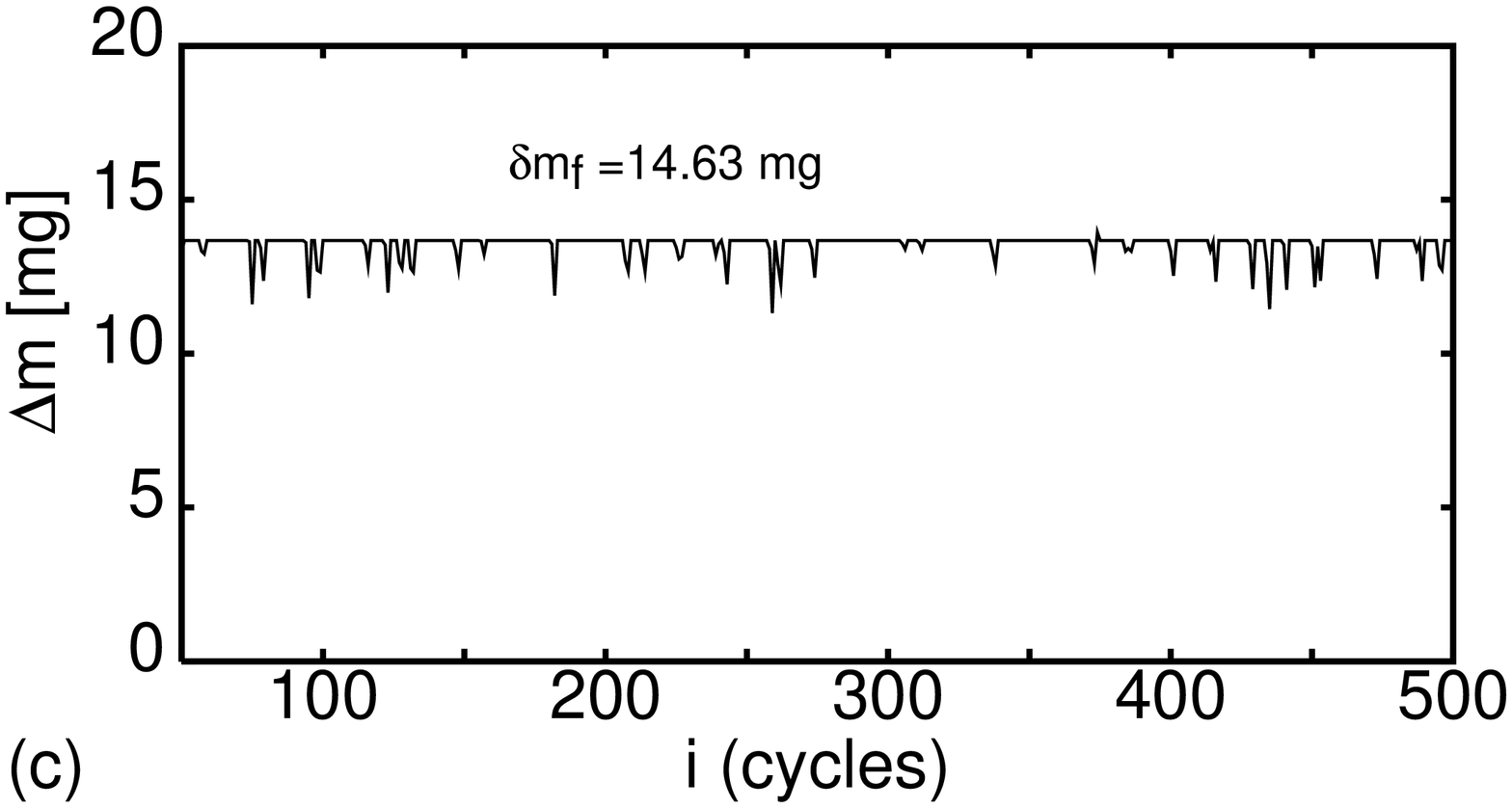}
\includegraphics[scale=0.3,angle=-90]{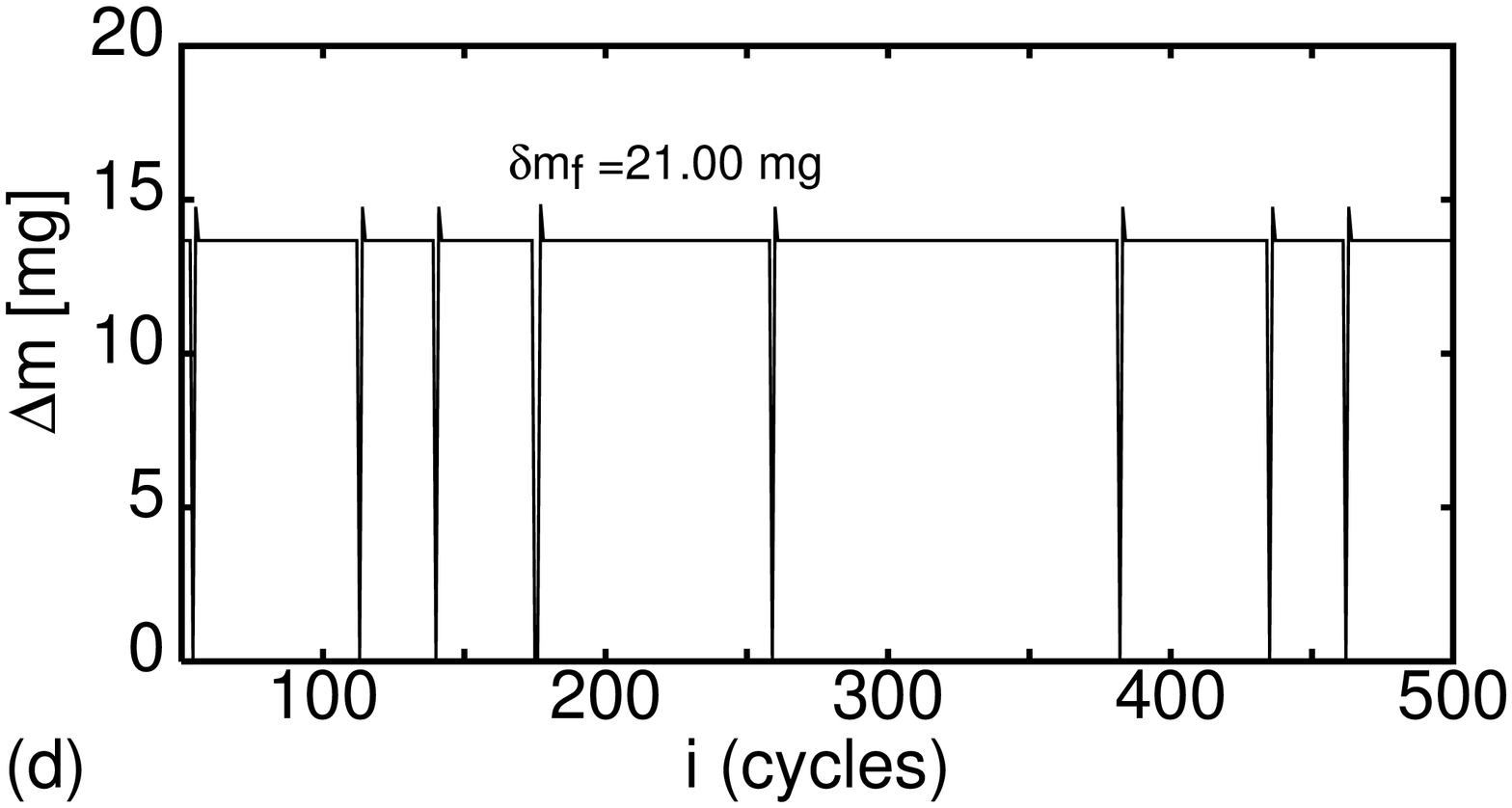}
\caption{The dependence of burned fuel mass 
on sequential cycles $i$ for deterministic (a) and
stochastic (b-d) processes.
$\delta m_a=200$ mg while
$\delta m_f$ takes different values: 13.50 mg , 14.63 mg, 21.00 mg denoted in
particular figures a-c.}
 \end{center}
\end{figure} 

The results for burned fuel mass $\Delta m_f$ are presented in Fig. 3. 
Starting from   
deterministic conditions ($\delta m_{f}=\delta m_{fo}={\rm const.}$)
we obtain the constant fraction of the burned fuel mass $\Delta m_f$
represented by the three straight lines in Fig. 3a lying very close to each 
other.
In Fig 3 b-c we show the same, $\Delta m_f$, for the considered case of 
assumed 
fuel injection ($\delta m_{fo}=13.50$ mg - Fig. 3b,
$\delta m_{fo}=14.63$ mg - Fig. 3c,
$\delta m_{fo}=21.00$ mg - Fig. 3d)
 and stochastic conditions. Due to different magnitudes parameter $\lambda$
fluctuations, and dependence of combustion curve Fig. 1 it is not 
surprise that the fluctuations of $\Delta m_f$ have different character in 
all these cases. For lean combustion, which is a stable process in 
deterministic case, the fuel injection fluctuations introduce considerable 
instabilities to the combustion process leading to the suppression of 
combustion because in some cycles (Fig. 3b) where $\lambda$ is larger that 
$1.3$. Then Equation  2.7 is not satisfied. In the next case  (Fig. 3c) 
the 
effect 
of stochasticity 
is much smaller. Here we have optimal air-fuel mixture. 
First of all one should note that fluctuations of 
$\lambda$ are smaller than in previous case (Fig. 2b). Moreover  
$\lambda$
oscillate around  
the region ($\lambda 
\approx 1$)  in
combustion curve 
(Fig. 1) which does not have big changes comparing to previous case.
Finally, Fig. 3d shows the sequence of $\Delta m_f$ for the large
$\delta m_{fo}$ (rich fuel-air mixture). The fluctuations of $\lambda$ 
are 
the smallest of all 
three ones but $\lambda \approx 0.7$ causes suppressions of combustions
 in some cycles similarly to the case shown in Fig. 3b.

\section{Conclusions}

In this paper we examined the origin of combusted mass fluctuations.
In case of stochastic conditions we have shown that depending on the 
quality of fuel-air mixture the final effect is different. The worse
situation is for lean combustion. The consequences of it can be observed
for idle speed regime of engine work. Unstable engine work, interrupted by
the  cycles without combustion lead to a large increase of fuel use.

Although the presented two component model is very simple it can reflect
the underlying nature of engine working conditions. In spite of fact 
that the model is characterized by the nonlinear transform 
(Eqs. 2.3, 2.5 and 2.7) similar to logistic one, we have not 
found 
any chaotic region.
Possibly that such solutions can be found for non realistic model 
parameters like $\lambda$ and $\alpha$.
The other strong limitation was 
concerned with the sharp edges of combustion curve
($\Delta m_f$ versus $\lambda$ Fig.1) modelled by a Heaviside step 
$\Theta(x)$ 
function Eq. 2.7. We used such an approximation as  
 a simplest one  but modeling with the exponential growth $\exp(-1/x)$
is more 
realistic and possible. Similar assumptions of the exponential dependence 
led to 
chaotic behaviour in papers  (Daw {\em et al.} 1996, 1998, Wendeker 2003).
From a physical point of view mixture 
gasoline-air is 
not uniform before ignition and that can cause nonuniform combustion
smearing the edges of the combustion curve Fig. 1.        
Calculations considering this effect are in progress and the results will 
be reported in 
a separate future publication.

\section*{References}

\noindent 
Clerk D., 1886,  The gas engine, Longmans, Green \& Co.,
London. \\ \\
Daw, C.S., Finney, C.E.A., Green Jr.,J.B.,
Kennel,
M.B.,
Thomas J.F. and
Connolly F.T., 1996, 
A simple model for cyclic variations in a spark-ignition engine, SAE
Technical Paper No.  962086. \\ \\
Daw, C.S., Kennel, M.B., Finney 
C.E.A., Connolly F.T., 1998
Observing
and
modelling dynamics in an
internal combustion engine, Phys. Rev. E, 57,
2811--2819. \\ \\
Daw C.S., Finney C.E.A. and Kennel M.B., 2000, Symbolic
approach for 
measuring temporal "irreversibility",  Phys. Rev. E,
62,  1912--1921.  \\ \\
Heywood JB., 1988,  Internal combustion engine 
fundamentals,  McGraw-Hill, New
York. \\ \\
Hu Z., 1996, Nonliner instabilities of 
combustion 
processes and
cycle-to-cycle variations
in spark-ignition engines, SAE Technical Paper No. 961197. \\ \\
Kowalewicz A., 1984,  Combustion Systems of High-Speed Piston
I.C. Engines, Studies in Mechanical Science 3, Elsevier, Amsterdam. \\ \\
Roberts J.B., Peyton-Jones J.C. and Landsborough
K.J., 1997, Cylinder
pressure variations as a
stochastic process, SAE Technical Paper No. 970059. \\ \\
Wendeker M.,  Niewczas A. and Hawryluk B., 1999, A
stochastic model of
the fuel injection of the
si engine,  SAE Technical Paper No. 00P-172. \\ \\
Wendeker, M., Czarnigowski, J., Litak, G. and  
Szabelski, K., 2003, Chaotic
combustion
in spark ignition engines, Chaos, Solitons \& Fractals 18, 805--808. 
\\ \\
Wendeker, M., Litak, G., Czarnigowski, J., and 
Szabelski, K., 2004, Nonperiodic oscillations 
of pressure in a spark ignition
engine,  Int. J. Bifurcation and Chaos  14, in press. 
\end{document}